\documentclass[twocolumn]{aastex62}
\usepackage{amsmath}

\renewcommand{\sc}{{\rm sc}}

\renewcommand{\vec}[1]{\mathbf{#1}}

\newcommand{\aave}[1]{\left\langle #1\right\rangle}

\graphicspath{{./}{figures/}}

\shorttitle{Characteristics of SB and NSB turbulence observed by \emph{PSP} }
\shortauthors{Bourouaine et al.}

\begin{document}

\title{Turbulence characteristics of switchbacks and non-switchbacks intervals observed by \emph{Parker Solar Probe}}
\correspondingauthor{Sofiane Bourouaine}
\email{sofiane.bourouaine@jhuapl.edu}

\author[0000-0002-2358-6628]{Sofiane Bourouaine}
\affil{Department of Aerospace, Physics and Space Sciences, Florida Institute of Technology,\\ 150 W University Blvd, Melbourne, Fl, 32901, USA}

\affil{Johns Hopkins University, Applied Physics Laboratory, Laurel, MD 20723, USA}

\author[0000-0002-8841-6443]{Jean C. Perez}
\affil{Department of Aerospace, Physics and Space Sciences, Florida Institute of Technology,\\ 150 W University Blvd, Melbourne, Fl, 32901, USA}

\author[0000-0001-6038-1923]{Kristopher C. Klein}
\affil{Lunar and Planetary Laboratory, University of Arizona, Tucson, AZ 85719, USA}

\author[0000-0003-4529-3620]{Christopher H. K. Chen}
\affil{School of Physics and Astronomy, Queen Mary University of London, London E1 4NS, UK}

\author[0000-0003-4529-3620]{Mihailo Martinovi\'c}
\affil{Lunar and Planetary Laboratory, University of Arizona, Tucson, AZ 85719, USA}

\author[0000-0002-1989-3596]{Stuart D. Bale}
\affil{Space Sciences Laboratory, University of California, Berkeley, CA 94720, USA}

\author[0000-0002-7077-930X]{Justin C. Kasper}
\affil{Department of Climate and Space Sciences and Engineering, University of Michigan, Ann Arbor, MI 48109, USA}

\author[0000-0003-2409-3742]{Nour E. Raouafi}
\affil{Johns Hopkins University, Applied Physics Laboratory, Laurel, MD 20723, USA}
\begin{abstract}

We use \emph{Parker Solar Probe} (\emph{PSP}) in-situ measurements to analyze the characteristics of solar wind turbulence during the first solar encounter covering radial distances between $35.7R_\odot$ and $41.7R_\odot$.  In our analysis we isolate so-called switchback (SB) intervals (folded magnetic field lines) from non-switchback (NSB) intervals, which mainly follow the Parker spiral field. Using a technique based on conditioned correlation functions, we estimate the power spectra of Elsasser, magnetic and bulk velocity fields separately in the SB and NSB intervals. In comparing the turbulent energy spectra of the two types of intervals, we find the following characteristics: 1) The decorrelation length of the backward-propagating Elsasser field $z^-$ is larger in the NSB intervals than the one in the SB intervals; 2) the magnetic power spectrum in SB intervals is steeper, with spectral index close to -5/3, than in NSB intervals, which have a spectral index close to -3/2; 3) both SB and NSB turbulence are imbalanced with NSB having the largest cross-helicity, 4) the residual energy is larger in the SB intervals than in NSB, and 5) the analyzed fluctuations are dominated by Alfv\'enic fluctuations that are propagating in the \emph{sunward} (\emph{anti-sunward}) direction for the SB (NSB) turbulence. These observed features provide further evidence that the switchbacks observed by \emph{PSP} are associated with folded magnetic field lines giving insight into their turbulence nature.
\end{abstract}

\keywords{}

\section{Introduction} \label{sec:intro}

Near-sun solar wind observations by \emph{Parker Solar Probe} (\emph{PSP}) during the first perihelion passes have revealed the frequent presence of the so-called magnetic ``switchbacks'' (SBs), which refer to local reversals of the radial magnetic field \citep{bale19,kasper19,horbury20,mcmanus20,td20}. These switchbacks are usually associated with correlated enhancements of the radial plasma flow. The origin of these field reversals is still under investigation and it is unclear whether they arise from impulsive events in the lower corona~\citep{roberts18,tenerani20} or they form in situ due to, for example, Alfv\'enic turbulence in the expanding wind \citep{squire20}.

SBs have been previously observed in fast solar-wind streams near 0.3 au~\citep{horbury18} and near or beyond 1 au~\citep{kahler96,balogh99}.  However, the SBs observed recently by \emph{PSP} during its first perihelion are detected in slow solar-wind, and they are sharper and ubiquitous. These observed slow solar-wind streams are identified as originating from an equatorial coronal hole and are dominated by Alfv\'enic fluctuations  \citep{bale19,kasper19}. The correlation between the speed enhancement and the presence of field reversals has been studied using \emph{Ulysses} (beyond 1 au) \citep{matteini14}. This correlation is interpreted as a result of a geometrical effect of the propagating large-scale Alfv\'enic fluctuations.

One interesting feature that characterizes the electron plasma in SB intervals as observed recently using the Solar Probe Analyzers SPAN-e instrument \citep{whittelsey20} on \emph{PSP} is that the electron strahl pitch angle distributions follow the magnetic field through SBs \citep{kasper19}. \emph{Ulysses} measurements also show that the relative proton beam appears to move slower than the proton core (in the spacecraft frame) following the reversed local field \citep{neugebauer13}. Furthermore, \cite{yamauchi04} showed that if the magnetic field is folded back on itself, alpha particles will locally have radial flow speeds less than protons.  The relative motion of protons and alphas during switchback periods near 0.3 au has been also discussed in~\cite{matteini15}. A more recent study of proton core-beam reversals with \emph{PSP} data shows that the proton core parallel temperature is the same inside and outside of switchbacks~\citep{woolley2020}, indicating more evidence that Alfv\'enic pulses travel along open magnetic field lines.

Another important  observed feature of SBs is that turbulence is dominated by sunward-propagating Alfv\'enic fluctuations over anti-sunward ones, an indication that Alfv\'enic fluctuations propagating away from the sun are following folded magnetic field lines. This feature has been observed near the sun at 0.16 au  using \emph{PSP} \citep{mcmanus20}, and also at heliocentric distances larger than 1 au using \emph{Ulysses} \citep{balogh99}. 

In this paper we investigate turbulence characteristics in the SB and NSB regions using a technique based on conditioned correlation functions. In section~\ref{sec:data} we present our data analysis and methodology to identify SB and NSB time intervals. In section~\ref{sec:turblence} we estimate the correlation functions, the power spectra, the normalized cross-helicity and the residual energy corresponding to those time intervals. Finally, in section~\ref{sec:conclusions} we conclude and discuss our results.

\section{Data and Methodology\label{sec:data}}

In our analysis we investigate the SBs that were observed during the first encounter (E1) between November 3 and November 9 of 2018. During this time period \emph{PSP} observed heliocentric distances ranging from $35.7$ $R_\odot$ to about $41.7$ $R_\odot$ \citep{fox16}. We use combined plasma and magnetic field measurements in the level-three data from SWEAP and level-two data from FIELDS on board \emph{PSP}, respectively. Data were obtained from the Solar Probe Cup (SPC)~\citep{case20} and the FIELDS flux-gate magnetometer (MAG)~\citep{bale16}. During this period, the data were measured with a time resolution of about 0.874~s for plasma moments and 0.22~s for the magnetic field vector. In our study we fit plasma and averaged magnetic field data to an uniform time grid with the resolution of the plasma data (i.e., 0.874~s). We also use a time-domain Hampel filter to remove artificial spikes within the plasma data~\citep{liu04}. 

\begin{figure}[!t]
    \centering
    \includegraphics[width=0.5\textwidth]{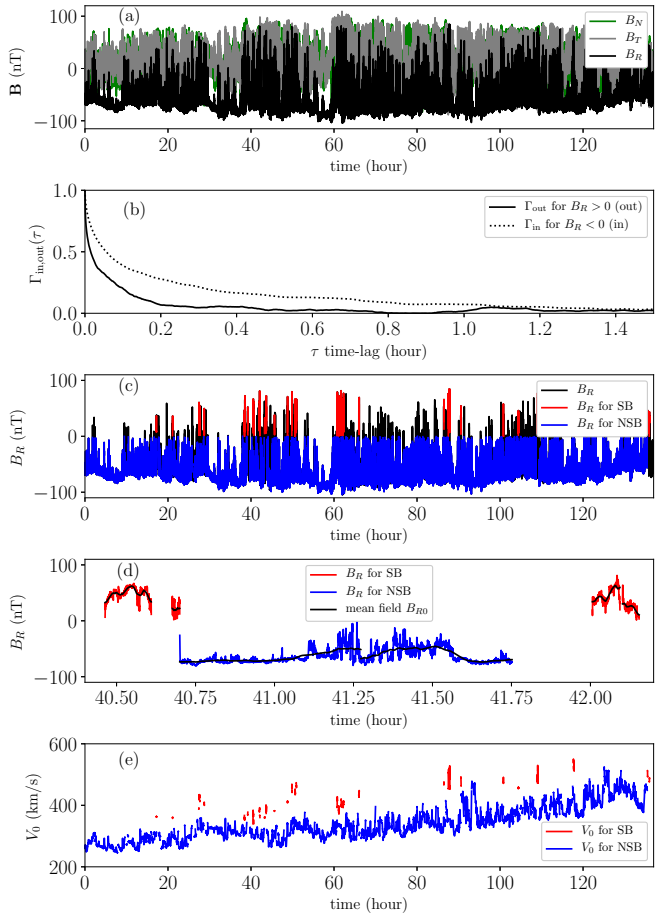}
     \vspace*{-0.1cm}
    \caption{Panels from top to bottom correspond to \emph{PSP} observations from November 3 2018 to November 9 2018. (a) time signal for magnetic field vector in RTN coordinates, (b) the normalized correlation of the magnetic field conditioned with $B_{\rm R}>0$ and $B_{\rm R}<0$ plotted vs. time lag $\tau$, (c) the radial component $B_{\rm R}$ that belongs to the selected to SB intervals (red), NSB intervals (blue) and non-selected intervals (black), (d) for clarity we plot a subset of isolated SB (red) and NSB (blue) intervals within a two-hour long subinterval, and (e) the mean velocity $V_0$ for SB (red) and NSB (blue) intervals.}
    \label{fig:fig1}
\end{figure}

Figure~\ref{fig:fig1}(a) shows the three components of the magnetic field vector in the RTN coordinate system  (where R is the direction from the sun's center to the spacecraft, T is the tangential component that results from the cross-product of the solar rotation vector with R, and N is the normal component  that completes the right-handed system). The $B_{\rm R}$ component frequently flips from sunward to anti-sunward direction during some time intervals. Overall, the magnetic field is strongly deflected from the expected Parker field line with large angles. For a detailed description of these time signals see~\cite{td20}.

In this paper we isolate the intervals that correspond to SBs (NSBs) within regions where $B_{\rm R}>0$ ($B_{\rm R}<0$). 
Before we isolate the SB and the NSB intervals we first estimate the decorrelation times $\tau_{\rm c, out}$ and $\tau_{\rm c, in}$ of magnetic field fluctuations in regions where the magnetic field radial component points outward (from the sun) $B_{\rm R}>0$ and inward $B_{\rm R}<0$, respectively. We estimate those decorrelation times for the magnetic field using the following conditioned correlation functions,  
\begin{eqnarray}
C_{\rm out}(\tau)&=&\aave{\left(\vec B(t)-\overline{\vec B}\right)\cdot \left(\vec B(t+\tau)-\overline{\vec B}\right)}_{B_{\rm R}>0}, 
\label{Eq.Corrvper}\\
C_{\rm in}(\tau)&=&\aave{\left(\vec B(t)-\overline{\vec B}\right)\cdot \left(\vec B(t+\tau)-\overline{\vec B}\right)}_{B_{\rm R}<0}, 
\label{Eq.Corrb}
\end{eqnarray}
where the symbols  $\aave{\cdots}_{B_{\rm R}>0}$ and $\aave{\cdots}_{B_{\rm R}<0}$ denote the ensemble average, which can be computed over many realizations (or averaged over time $t$) conditioned by $B_{\rm R}>0$ or $B_{\rm R}<0$. Here $\overline{\vec B}$ are obtained by averaging over the same ensemble, i.e., all those points used in the calculation of the correlation functions. The time lag $\tau$ is varied from 0 to 1.5 hour.

 Figure~\ref{fig:fig1}(b) displays the normalized correlation functions $\Gamma_{\rm out}(\tau)\equiv C_{\rm out}(\tau)/C_{\rm out}(0)$ (solid line) and $\Gamma_{\rm in}(\tau)\equiv C_{\rm in}(\tau)/C_{\rm in}(0)$ (dotted line) as a function of $\tau$. We find that the magnetic decorrelation times are $\tau_{\rm c,in}\simeq 2$~min and $\tau_{\rm c,out}\simeq 6$~min (times for which $\Gamma_{\rm out,in}=1/e$), meaning that the magnetic turbulent fluctuations in strongly reversed field decorrelate faster in time. Note that the obtained decorrelation time $\tau_{\rm c,out}\simeq 6$~min is within the variation range of the magnetic decorrelation time found by~\cite{chen20,parashar20}. 
 
We next identify the SB (NSB) intervals inside the $B_{\rm R}>0$ ($B_{\rm R}<0$) regions. One criterion we use to isolate the SB and NSB intervals is to consider continuous measurements of $B_{\rm R}>0$ ($B_{\rm R}<0$) that last at least $T\gtrsim \tau_{\rm c,out}$ ($\tau_{\rm c,in}$) for SB (NSB) intervals. In this way we designate SB and NSB time intervals with time lengths equal or higher than the corresponding magnetic decorrelation time periods, allowing for a proper estimation of the correlated fluctuations and their associated local mean field.

In Figure~\ref{fig:fig1}(c) we plot the radial component $B_{\rm R}$ associated with the SB (red) and NSB (blue) time intervals. The intervals with black colors are those intervals that do not satisfy the criteria described above. Figure~\ref{fig:fig1}(d) shows the SB and NSB intervals within a two-hour sub-interval to illustrate this selection process.

We use the obtained decorrelation times, $\tau_{\rm c,out}$ and $\tau_{\rm c,in}$ to define the local mean magnetic field within the SB and NSB intervals as 
\begin{equation}
\vec B_0(t_j)= \frac 1{N_j}\sum_{i}W_T(t_j-t_i)\vec B(t_i),
\label{Eq.mean_field_NSB}
\end{equation}
where $N_j$ is the number of averaging samples and $W_T(t)$ is a windowing function that vanishes everywhere except for $|t|\le T/2$ where it is equal to one. The period $T$ is chosen to be equal to the decorrelation time such that $T=\tau_{\rm c,out}$ (for SBs) or $T=\tau_{\rm c,in}$ (for NSBs). The radial component $B_{0{\rm R}}$ are plotted in black in Figure~\ref{fig:fig1}(d). The mean velocity $\vec V_0$ can be estimated in a similar fashion by replacing $\vec B$ with $\vec V$ in Eq.~\eqref{Eq.mean_field_NSB}. Figure~\ref{fig:fig1}(e) displays the mean speed $V_0$ for the SB (red) and NSB (blue) intervals. The Figure shows that the mean velocity profiles for SB and NSB have an increasing trend from values 340 km/s to 550 km/s for SB and from 245 km/s to 520 km/s for NSB. The percentage increase of the mean velocity in the NSB intervals is about $100\%$ vs a 60$\%$ increase in SB intervals. 

Determining the local mean magnetic field allows for the estimation of the local angle $\theta_b\equiv (\hat{\vec R},\vec B_0)$ between the radial direction $\hat{\vec R}$ and the local magnetic field.

\begin{figure}
    \centering
    \includegraphics[width=0.47\textwidth]{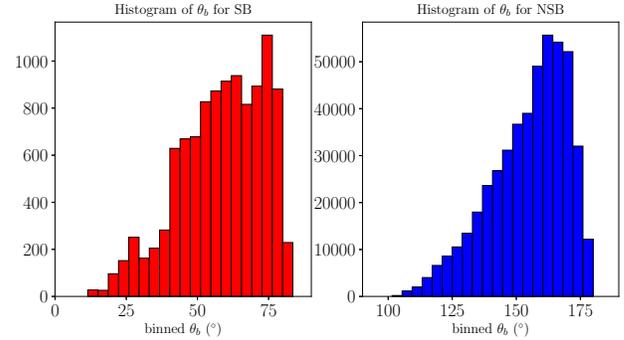}
       \vspace*{0.cm}
    \caption{Histograms illustrating the number of points that correspond to the angle $\theta_b$ in the SB (left panel) and NSB (right panel) intervals.}
     \label{fig:fig2}
\end{figure}

Figure~\ref{fig:fig2} displays the histogram for the occurrence of the angle $\theta_b$ for SB configurations (left panel) and NSB configurations (right panel). As expected, the most common angle $\theta_b$ for NSB intervals is $\theta_b\simeq 170^\circ$ which is close to the Parker field angle. The angle $\theta_b$ varies from 100$^\circ$ to 175$^\circ$ in the NSB intervals, and from about  10$^\circ$ to 80$^\circ$ in the SB intervals. The most likely angles $\theta_b$ in SB  intervals range between 40$^\circ$ and 75$^\circ$.

It is worth mentioning that the outcomes from the analysis below are not sensitive to the choice of $\theta_b$ range for SB and NSB intervals. For example, we repeated the analysis and considered a narrow range of $\theta_b$ with $\theta_b>150^\circ$ characterizing NSB intervals and a wide range of $\theta_b$ with $\theta_b<150^\circ$ characterizing SB intervals, and the obtained results were very comparable the ones presented below.

\section{Turbulence characteristics in SB and NSB intervals \label{sec:turblence}}

We use a technique based on so-called conditioned correlation functions (described in \cite{bourouaine20}) to estimate the power spectra from discontinuous intervals. The power spectra that correspond to Elsasser fields $\vec z^\pm$, velocity field $\vec V$ and magnetic field $\vec B$ from these discontinuous SB and NSB intervals can be then derived from their corresponding conditioned correlation functions. Here, the Elsasser fields are defined as $\vec z^\pm=\vec V \pm \vec V_A$, where $\vec V_A=\vec B/\sqrt{4\pi\rho}$ is the Alfv\'en velocity where $\rho$ is the proton mass density. 

In this technique we treat these discontinuous intervals as statistical realizations to estimate various correlation functions, and simply create ensembles in which the fields have certain properties (conditioning) under an implicit underlying assumption which is that ergodicity applies to these ensembles. The Fourier transform of these correlation functions can be used to determine the corresponding power spectra. Let us first define the conditioned correlation functions of the various fields for SB and NSB as follows 
\begin{eqnarray}
C^{\rm SB}_q(\tau)&=&\aave{\left(\vec q(t)-\overline{\vec q}\right)\cdot \left(\vec q(t+\tau)-\overline{\vec q}\right)}_{{\rm SB}, V_0}, 
\label{Eq.Corrsb}\\
C^{\rm NSB}_q(\tau)&=&\aave{\left(\vec q(t)-\overline{\vec q}\right)\cdot \left(\vec q(t+\tau)-\overline{\vec q}\right)}_{{\rm NSB}, V_0},
\label{Eq.Corrnsb}
\end{eqnarray}
where the generic vector $\vec q$ can represent the Elsasser variables $\vec z^\pm$, the velocity $\vec V$ or magnetic $\vec B$ fields. The correlation functions are computed over many realizations (or averaged over time $t$) conditioned by SB or NSB selected intervals (defined above) and the solar wind speed $V_0$. Here, we condition the correlations by considering only the statistics of those two times $t$ and $t+\tau$ in Equations~\eqref{Eq.Corrsb}~and~\eqref{Eq.Corrnsb} that belong to SB and NSB intervals, respectively. In order to ensure that we are dealing with the same stream or turbulence we do not allow for a significant variation in the mean solar-wind velocity. Therefore we condition the correlation functions with respect to mean velocity $V_0$ and consider only those points where their corresponding values of $V_0$ vary from a minimum value $V_{0,{\rm min}}$ corresponding to SB or NSB intervals to a maximum value $V_{0,{\rm max}}\simeq 1.5\times V_{0,{\rm min}}$ (a 50$\%$ increase).

There are three main advantages of calculating the power spectra through the correlation functions: 1) one can deal with discontinuous intervals (such as the SB and NSB intervals for our case) and avoid any unwanted points, including gaps of bad measurements in the calculation of the correlation functions, 2) one can obtain the correlation functions at time-lags (including correlation times) much longer than the periods of the used discontinuous intervals, and 3) one can check the statistics that correspond to the estimation of the correlation functions for each time-lag $\tau$. In addition, further conditioning the correlation functions, for example with respect to solar wind velocity, ensures that discontinuous intervals are associated with a single stream (or single ``kind of'' turbulence). In many cases, using the standard method of estimating the power spectra through the Fourier transform of continuous time signals does not guarantee one can avoid mixing measurements from qualitatively different streams, such as when the spacecraft is crossing solar wind streams with a significant difference in speed. Such continuous time signals mixing different speeds may artificially increase the kinetic energy of the outer-scale.

\begin{figure}
    \centering
    \includegraphics[width=0.5\textwidth]{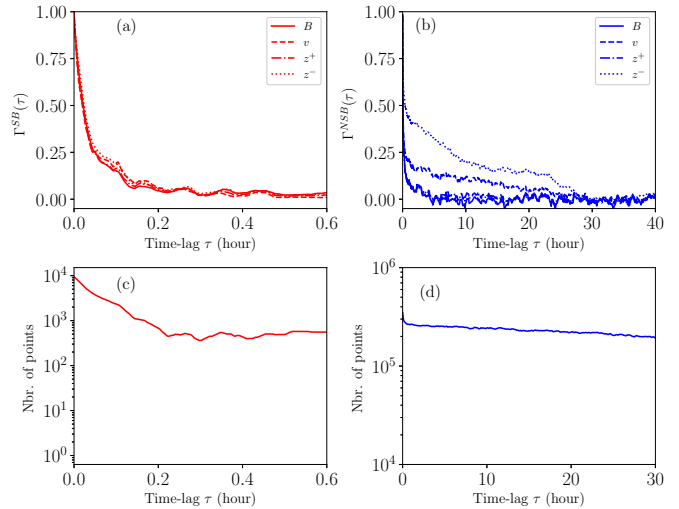}
       \vspace*{-0.5cm}
    \caption{ 
    Upper panels: the normalized correlation function for the variables $q$ using for SB intervals (panel (a)) and for NSB intervals (panel (b)). All plotted versus time-lag $\tau$. Lower panels: the number of points that correspond to the calculation of the correlation function for the SB (panel (c)) and NSB (panel (d)) intervals. The SB correlation functions drop to zero at time-lag that is much smaller than 0.6 hour, and no need to extend the time-lag to values bigger than 0.6 hour where the number of statistical points starts to decrease.}
    \label{fig:fig3}
\end{figure}

Figure~\ref{fig:fig3}(a)~and~\ref{fig:fig3}(b) show the normalized correlation functions 
$\Gamma^{\rm SB}_q(\tau)\equiv C^{\rm SB}_{q}(\tau)/C^{\rm SB}_{q}(0)$ and $\Gamma^{\rm NSB}_q(\tau)\equiv C^{\rm NSB}_{q}(\tau)/C^{\rm NSB}_{q}(0)$ computed using equations (\ref{Eq.Corrsb}) and (\ref{Eq.Corrnsb}), respectively. Interestingly enough, the curves of the normalized correlation functions that correspond to SB intervals drop relatively faster than those ones corresponding to NSB intervals. For NSB intervals, the largest decorrelation time corresponds to the Elsasser field $\vec z^{-}$, which is about $\simeq 3$ hours, then after, the velocity decorrelation time is about $\simeq 9$ min. Also, the decorrelation times corresponding to Elsasser field $\vec z^{+}$ and magnetic field $\vec B$ are identical and close to $5$ min in the NSB intervals. However, for SB intervals, the decorrelation times corresponding to all fields $q$ seem to be identical and close to 2 min. 

 To estimate the corresponding perpendicular correlation lengths from the obtained correlation times we invoke Taylor's Hypothesis. Although the ratio $V_A/V_\perp \simeq 1$ (where $V_\perp$ is the perpendicular velocity of the spacecraft seen in the plasma frame),  \citet{bourouaine19,bourouaine20,perez20} showed that Taylor approximation can still be valid if the dimensionless parameter $\epsilon=\delta u_0/(\sqrt{2} V_\perp) \lesssim 0.5$ (where $\delta u_0$ is the outer-scale fluid velocity) and under the assumption that the turbulence is strong. Here we found that $\epsilon\simeq 0.15\pm0.02$ ($\epsilon\simeq 0.30\pm0.08$) for SB (NSB) intervals. $V_\perp$ is estimated by taking into account the spacecraft velocity, $V_{\rm sc}=92$ km/s that is tangential. The errors in $\epsilon$ are estimated due to error in velocity fluctuation caused by the noise and the uncertainty in the bulk velocity. The correlation lengths can be obtained through the decorrelation time $t_c$ as $\lambda_c\simeq V_\perp t_c$, where $V_\perp=253$ km/s ($V_\perp=119$ km/s) for SB (NSB) intervals. This leads to correlation lengths  $\lambda_c\simeq 30,000$ km for $B$ , $z^\pm$ and $V$ fields in SB intervals. In NSB intervals, the correlations length of $B$ and $z^+$ is $\lambda_c\simeq 35,000$ km, which is comparable to the one in SB intervals, and the correlation lengths corresponding to $V$ and $z^-$ fields are about 64000 km and 1 Mm, respectively.

Figures~\ref{fig:fig3}(c)-(d) show the number of points that corresponds to the calculations of the correlation functions in SB (lower left) and NSB (lower right) intervals. As expected, we have a smaller sample sizes for SB intervals. However, these correlation functions were still computed with a sufficient statistics for time-lag $\tau \le 0.6$ hour that is much higher than the correlation times in the SB intervals.
 
\begin{figure}
    \centering
    \includegraphics[width=0.5\textwidth]{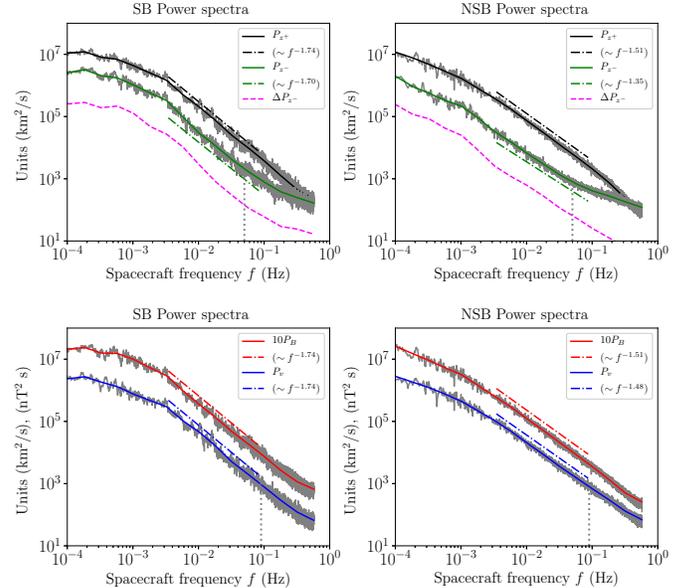}
       \vspace*{-0.5cm}
    \caption{ Left panels: Spacecraft-frame power spectra of $\vec v,\vec b$ and $\vec z^\pm$ for SB intervals. Right panels: Spacecraft-frame power spectra $\vec v,\vec b$ and $\vec z^\pm$ for NSB intervals. All spectra are plotted versus frequency $f$. Power-law fits with spectral index (colored dash-dotted lines) are added. The magnetic power spectra were multiplied by a factor of 10 for easy viewing. The vertical dotted line were plotted to refer to the noise level in the power spectra for $\vec v$ and $\vec z^{-}$. All the the power spectra were multiplied by factor of 2 to account for their negative frequency part.  The uncertainty on $z^-$ power spectra were plotted with dashed magenta lines.}
    \label{fig:fig4}
\end{figure}

In Figure \ref{fig:fig4} we plot the power spectra of all variables $\vec q$ and that  were computed as the Fourier transform of their corresponding correlation functions (plotted in Figure \ref{fig:fig3}) 
\begin{equation}
P_q(f)\equiv\frac 1{2\pi}\int_{-\infty}^\infty C_q(\tau)e^{i2\pi f\tau}d\tau. 
\end{equation}
The power spectra corresponding to SB (left panel) and NSB (right panel) intervals are plotted versus the frequency $f$ in  spacecraft frame. These power spectra were computed using correlation functions that span over time-lags that are much larger than the decorrelation times.

Our analysis shows that the energy of fluctuations in the SB intervals follows a power-law turbulent spectrum, indicating that a broad turbulence spectrum of fluctuations occurs within SB intervals.
Furthermore, it seems that the power spectra in SB intervals are relatively steeper following a power-law with spectral index of about -1.7 that is close to -5/3 for all variables, whereas in the NSB intervals the power spectra corresponding to $\vec B$, $\vec z^+$ and $\vec v$ follow a power-law with spectral index that is close to -3/2. However, the power spectrum corresponding to $\vec z^{-}$ in NSB follow a power-law with a spectral index -1.35. The noise levels in the velocity and the $\vec z^{-}$ power spectra are shown using two vertical dotted lines in Figure \ref{fig:fig4}. The noise level starts to occur at frequencies higher than $f\simeq 10^{-1}$ Hz in the velocity spectra, and higher than $\simeq 5\times 10^{-2}$ Hz in the $\vec z^-$ spectra. The estimation of the noise level were based on the high-frequency flattening observed in the velocity power spectrum calculated from continuous and evenly spaced seven-hour long time interval on November 6 (2018). In this time interval only a few interpolations were applied due to some missing points. The observed flattening in $\vec v$ power spectrum occurs at frequency $f\simeq 10^{-1}$ Hz that is consistent with one found in \cite{chen20}.

\begin{figure}
    \centering
    \includegraphics[width=0.5\textwidth]{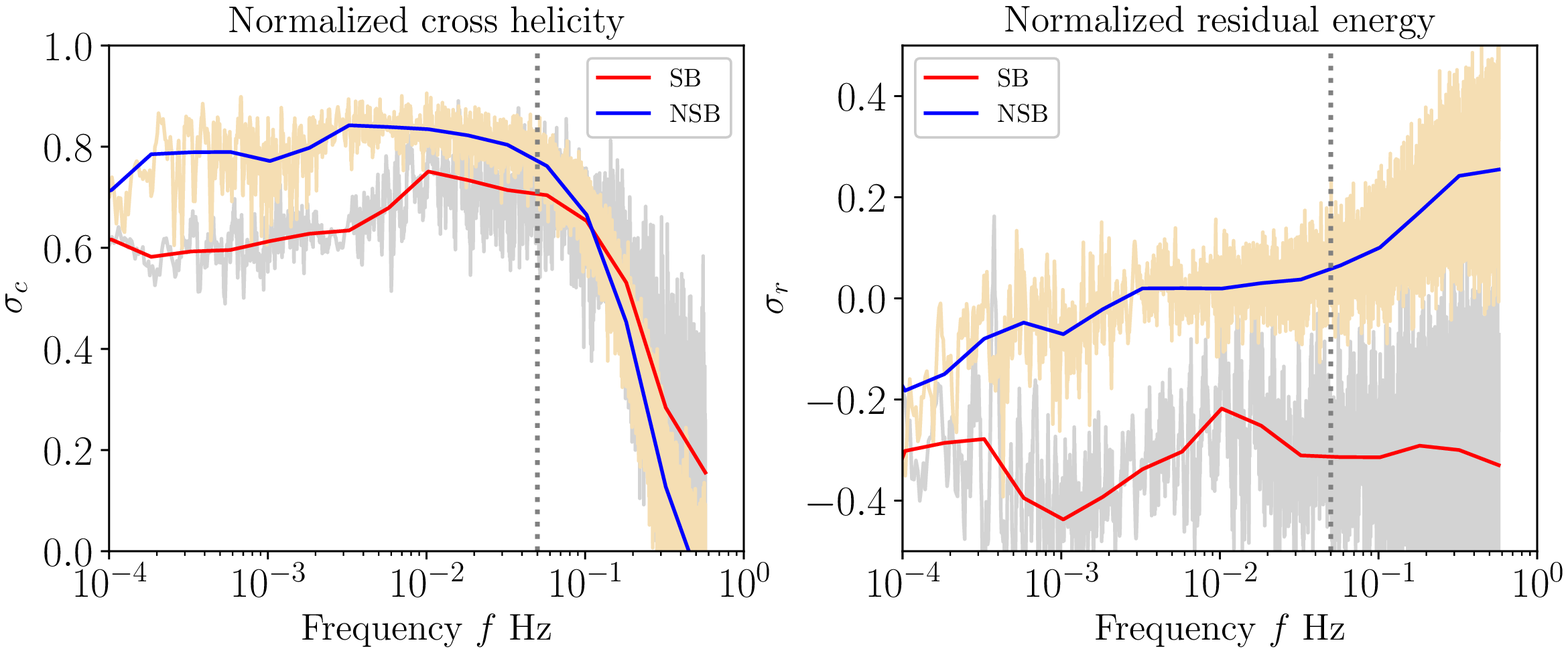}
       \vspace*{-0.5cm}
    \caption{ Left panel: Normalized cross-helicity for SB (red) and NSB (blue). right panel: Normalized residual energy for SB (red) and NSB (blue). All plotted as a function of the frequency $f$. The curves at frequencies higher than $5\times 10^{-2}$ Hz (vertical dotted line) may be contaminated by the noise.}
    \label{fig:fig5}
\end{figure}

  To rule out that the measured minor $z^-$ spectra are not spurious due to measurements uncertainties, we estimate their corresponding uncertainties for SB and NSB intervals in Figure \ref{fig:fig5} (magenta dashed lines) based on the uncertainties on velocity components and density. We used the upper estimate of the uncertainty on the moments due to, for example, the amount of background noise in the current spectra, alpha-proton VDF separation, etc., provided in the SPC L3I data. The averaged relative uncertainties on the velocity components are about $5\%$ for the radial component and about $20\%$ for the other components, and the relative uncertainty on the density is about $20\%$. The relative uncertainty on the $\vec z^-$ power spectrum is found to be about $10\%$ which indicates that the measured z- spectrum is physical and not spurious.

In Figure \ref{fig:fig5} we plot the normalized cross-helicity from the Elsasser fields as $\sigma_c=(P_{z^{+}}-P_{z^{-}})/(P_{z^{+}}+P_{z^{-}})$  (where $P_{z^{\pm}}$ is the power spectra corresponding to $\vec z^\pm$) and the normalized residual-energy $\sigma_r=(P_v-P_B)/(P_v+P_B)$ (where $P_v$ and $P_B$ are the power spectra corresponding to $\vec{v}$ and $\vec B$) for the SB (red) and NSB (blue) turbulence. The noise may affect the curves of $\sigma_r$ and $\sigma_c$ at frequencies higher than $5\times 10^{-2}$ Hz, and therefore, we exclude this frequency range from any physical interpretation.

Here we notice that the normalized cross-helicity of the turbulent field is positive for both SB and NSB intervals. The turbulence is then dominated by the $\vec z^{+}$ fluctuations, which means that the Alfv\'en fluctuations propagate anti-sunward in the NSBs and in sunward direction in the SBs. Furthermore, we found that the turbulence is more imbalanced in the NSB intervals than in the SB intervals.  

It is worth-noting that the magnitude of the normalized residual energy in the SB turbulence is higher than the one corresponding to NSB turbulence. Also, steepness in the magnetic power spectrum in the SB intervals may be due to the fact $|\sigma_r|$ for the SB turbulence is higher t than $|\sigma_r|$ for the NSB turbulence. The connection between the steepness of the magnetic spectrum and the normalized residual energy was studied in many previous works near 1 au \citep{boldyrev11}. The finding of the magnetic field and total energy spectra becoming steeper as the turbulence becomes more balanced is consistent with 1 au observations \citep{chen13a}.

\section{Discussion and Conclusion\label{sec:conclusions}} 

In this work we have used a methodology based on conditioned correlation functions to analyze the characteristics of the turbulence in the SB and NSB intervals at heliocentric distances from $35.7R_\odot$ to $41.7R_\odot$. The power spectra that correspond to Elsasser fields as well as the velocity and magnetic fields were then computed as the Fourier transform of these conditioned correlations functions. We found that the correlation lengths of the variables $B$ and $z^+$ in SB and NSB intervals are comparable. However, the estimated correlation length of $z^-$ in NSB intervals is about 1 Mm and it is much larger that the one (about 30,000 km) in the SB intervals. The long correlation lengths (or correlation times) of $z^-$ within the NSB intervals are similar to those found by \cite{chen20}, which was interpreted as consistent with the generation of $z^-$ fluctuations from non-WKB (Wentzel–Kramers–Brillouin)  reflections of $z^+$ due to inhomogeneity of the background plasma along the magnetic field line~\citep{heinemann80,hollweg81,velli93,hollweg07,verdini12,chandran09,perez13,chandran19}. The new finding here is that in the SB intervals the correlation time is shorter and the same in all fields illustrates the different nature of the fluctuations within the switchbacks. Here, we conjecture that because non-WKB reflections occur more efficiently for AW fluctuations with long parallel wavelengths, of the order of one solar wind radius (as our estimates indicate in the NSB)~\citep{perez13,chandran19}, long-wavelength correlated fluctuations arising from reflections are suppressed within the SBs, as they are likely of much smaller extent.

Furthermore, we found that turbulence in SBs is more balanced than in NSBs, and the magnetic power spectrum corresponding to SB turbulence is steeper than that corresponding to NSB turbulence. This observed steepening in the power spectra with more balanced turbulence is also associated with larger amount of residual energy. This finding seems to be well consistent with previous 1 au measurements \citep{chen13a,bowen18}.  Moreover, one possibility for why the turbulence is more balanced in SB intervals is that it is locally driven, either by velocity shear introduced by the SBs or by the oppositely-directed waves on neighbouring field lines that are travelling in both directions due to the folded field.  Therefore, the analysis presented here contributes to a better understanding of the nature of the switchback turbulence and may provide empirical insights on the process of reflection-driven turbulence inside and outside switchbacks.

\acknowledgments
 SB was supported by NASA grants NNX16AH92G, 80NSSC19K0275 and 80NSSC19K1390. JCP was partially supported by NASA grants NNX16AH92G, 80NSSC19K0275 and NSF grant AGS-1752827.  K.G.K. is supported by NASA contract NNN06AA01C and 80NSSC19K0912. C.H.K.C. is supported by STFC Ernest Rutherford Fellowship ST/N003748/2 and STFC Consolidated Grant ST/T00018X/1. M.M is supported by NASA grant 80NSSC19K1390. We acknowledge the NASA Parker Solar Probe Mission, the SWEAP team led by J. Kasper, and the Fields team lead by S. Bale for use of data.

\end{document}